\documentclass[runningheads]{llncs}

\usepackage[T1]{fontenc}
\usepackage{amsmath,amssymb}
\usepackage{array}
\usepackage{booktabs}
\usepackage{graphicx}
\usepackage{url}
\usepackage{float} 
\usepackage[hidelinks]{hyperref}
\usepackage{orcidlink}
\renewcommand{\orcidID}[1]{\orcidlink{#1}}
\hypersetup{
    pdftitle={Toward Practical Decentralized Proof-of-Location via Physical Witnessing Zones},
    pdfauthor={Tamor Tomson, Eduardo Brito, Amnir Hadachi, Ulrich Norbisrath},
    pdfkeywords={Decentralized Proof-of-Location, Digital Trust Infrastructure, Location Verification}
}

\begin{document}

\newcommand\submittedtext{%
\textbf{Preprint}. This work has been accepted to the CBI \& EDOC 2026 Forum. Personal use of this material is permitted. Permission must be obtained for all other uses, in any current or future media.}

\newcommand\submittednotice{%
\begingroup
\renewcommand\thefootnote{}%
\footnotetext{%
\noindent\fbox{\parbox{\dimexpr\linewidth-2\fboxsep-2\fboxrule\relax}{\submittedtext}}%
}%
\endgroup
}

\title{Toward Practical Decentralized Proof-of-Location via Physical Witnessing Zones\thanks{This work was supported by the Charlemagne Prize Academy Fellowship and the Inclusive Technology Foundation (INTF), formerly XRPLF.}}
\titlerunning{Practical Decentralized Proof-of-Location}

\author{Tamor Tomson\inst{1} \and Eduardo Brito\inst{1,2}\orcidID{0009-0002-9996-6333} \and Amnir Hadachi\inst{1}\orcidID{0000-0001-9257-3858} \and Ulrich Norbisrath\inst{1}\orcidID{0000-0002-6151-991X}}
\authorrunning{T. Tomson et al.}

\institute{Institute of Computer Science, University of Tartu, Tartu, Estonia\\
\email{\{tamor.tomson,hadachi,ulrich.norbisrath\}@ut.ee}
\and
Cybernetica AS, Tallinn, Estonia\\
\email{eduardo.brito@cyber.ee}}

\maketitle
\submittednotice

\begin{abstract}
Digital services increasingly rely on claims that a person, device, or asset was in a specific place at a specific time. Today, those claims often depend on self-reported location data, which is easy to falsify and difficult to verify after the fact. Proof-of-Location (PoL) systems address this gap by turning presence claims into evidence that an independent verifier can later inspect. This paper builds upon recent theoretical work on decentralized PoL architectures and demonstrates how they can move from emulation to a physical prototype built with low-cost hardware. We implement a witnessing zone in which fixed nearby devices measure a prover's presence, exchange claims over a local mesh, and record them in a tamper-evident ledger. Building the prototype required adapting the abstract protocol to physical constraints through witness-initiated ranging, cross-witness consistency checks, and freshness binding against replay. Our controlled indoor evaluation shows that the system can produce accurate, low-latency proof objects while detecting simulated replay and malicious-ranging attacks. The result is a reusable experimental baseline for next-generation digital trust infrastructure, that exposes the remaining calibration, verifier-independence, radio-integrity, and scaling requirements for decentralized location evidence, where physical presence claims can be independently checked under real radio, networking, and timing conditions.

\keywords{Decentralized Proof-of-Location \and Digital Trust Infrastructure \and Location Verification}
\end{abstract}

\section{Introduction}
Many digital services depend on the assumption that a user, device, or digital artifact was physically present at a claimed place and time. Examples include geofenced services, supply-chain verification, mobile voting, and digital-evidence authentication. Recent work on location-data provenance maps the risks created when such claims are weakly sourced or self-reported~\cite{brito2026locationrisks}. As generative AI lowers the cost of fabricating plausible media, verifiable place-and-time evidence becomes one way to strengthen claims about real-world provenance~\cite{brito2026capturetime}.

The dominant location source on consumer devices, the Global Navigation Satellite System (GNSS), was not designed for adversarial settings. GNSS coordinates are self-reported and can be spoofed with inexpensive equipment. Wi-Fi and cellular positioning share similar weaknesses when a service trusts reported coordinates. A Proof-of-Location (PoL) system addresses this gap by producing a transferable certificate that allows a third-party verifier to check that a prover was near a set of witnesses at a particular time~\cite{saroiu2009locationproofs,brito2025decentralized}.

PoL research has evolved from centralized trusted-location authorities to distributed and blockchain-backed designs with privacy protections~\cite{waters2003secure,luo2010veriplace,zhu2011applaus,akand2023privacy,brito2026taxonomy}. Recent work organizes this design space into a taxonomy and methodology for mapping application requirements to PoL architectures~\cite{brito2026taxonomy}, and proposes a decentralized witnessing-zone model in which witnesses use distance bounding protocols, a local mesh, and permissioned Byzantine fault-tolerant consensus to store signed claims~\cite{brito2025decentralized}. That model was validated through emulation; feasibility under real radios, embedded hardware, and physical ranging conditions remains underexplored.

This paper turns that architecture into a physical research platform built from commodity embedded devices and evaluates the full path from UWB ranging to blockchain claim storage and independent proof verification. The main contributions are:
\begin{itemize}
    \item an exploratory end-to-end physical implementation of a decentralized PoL witnessing zone using ESP32-DWM3000 UWB nodes, Raspberry Pi 4 witnesses, B.A.T.M.A.N.-Adv mesh networking, and GoQuorum IBFT consensus;
    \item three protocol adaptations needed for practical hardware: witness-initiated double-sided two-way ranging (DS-TWR), a global RMS residual check for multilateration consistency, and block-reference freshness checking;
    \item an experimental campaign of 340 trials under nominal, mobility, range-manipulation, stale-reference, and out-of-zone conditions;
    \item empirical evidence that sub-meter decentralized PoL is feasible in the tested single-zone indoor setting on low-cost hardware, together with a research outlook on residual-threshold selection, time synchronization, mobile operation, privacy, and deployment beyond a single witnessing zone.
\end{itemize}

The novelty claimed here is the physical integration and evaluation of the previously emulated witnessing-zone model with real UWB radios, a Layer-2 mesh, and a running BFT ledger~\cite{brito2025decentralized}; we do not claim the constituent techniques as individually new.

\section{Background and Related Work}
\label{sec:background-related-work}
\subsection{Proof-of-Location}
A PoL protocol produces evidence that a prover was present at a claimed location at a particular time. Early work introduced location proofs for mobile applications~\cite{saroiu2009locationproofs} and secure private location evidence~\cite{waters2003secure}. Subsequent systems investigated privacy-aware architectures~\cite{luo2010veriplace}, witness-based proof collection~\cite{zhu2011applaus,wang2016stamp}, and collusion-resistant or geo-tamper-resistant proof generation~\cite{akand2023privacy,nosouhi2018sparse}. Recent work connects PoL to location-data provenance risks~\cite{brito2026locationrisks}, capture-time authenticity~\cite{brito2026capturetime}, and taxonomy-based system design~\cite{brito2026taxonomy}.

Secure positioning and distance bounding provide cryptographic and radio-layer tools for limiting how far a prover can be from a verifier~\cite{capkun2006secure,brands1994distance,rasmussen2010rf}. A transferable proof also needs evidence that an independent verifier can later check. Decentralized PoL systems therefore combine local witness evidence with tamper-evident storage and verification logic.

\subsection{Decentralized Witnessing Zones}
Some PoL designs store witness claims or proofs for later ledger audit~\cite{amoretti2018blockchain,nosouhi2020blockchain,wu2020zkpol}. The witnessing-zone architecture of Brito et al.~\cite{brito2025decentralized} organizes fixed witnesses into local fault-tolerant zones that discretize space and time. Each known-position witness contributes a signed distance attestation to a ledger, and the prover aggregates sufficient attestations. Unlike prior work demonstrating individual ranging, privacy, witness, or ledger mechanisms, we test whether the layers of this previously emulated architecture operate together on physical radios and embedded nodes.

The decentralized model distributes trust across three mechanisms rather than relying on a single location authority. First, several authenticated witnesses independently measure proximity to the prover. Second, only attestations produced within the same zone and time interval are eligible for proof construction, which binds location evidence to both space and time. Third, a quorum of witnesses commits accepted claims through Byzantine-fault-tolerant consensus, producing an ordered and tamper-evident claim log~\cite{brito2026capturetime}. In the common four-witness instantiation, a threshold of three attestations tolerates one faulty witness under the standard $n \ge 3f+1$ Byzantine fault-tolerance bound.

\begin{figure}[t]
\centering
\includegraphics[width=0.78\textwidth]{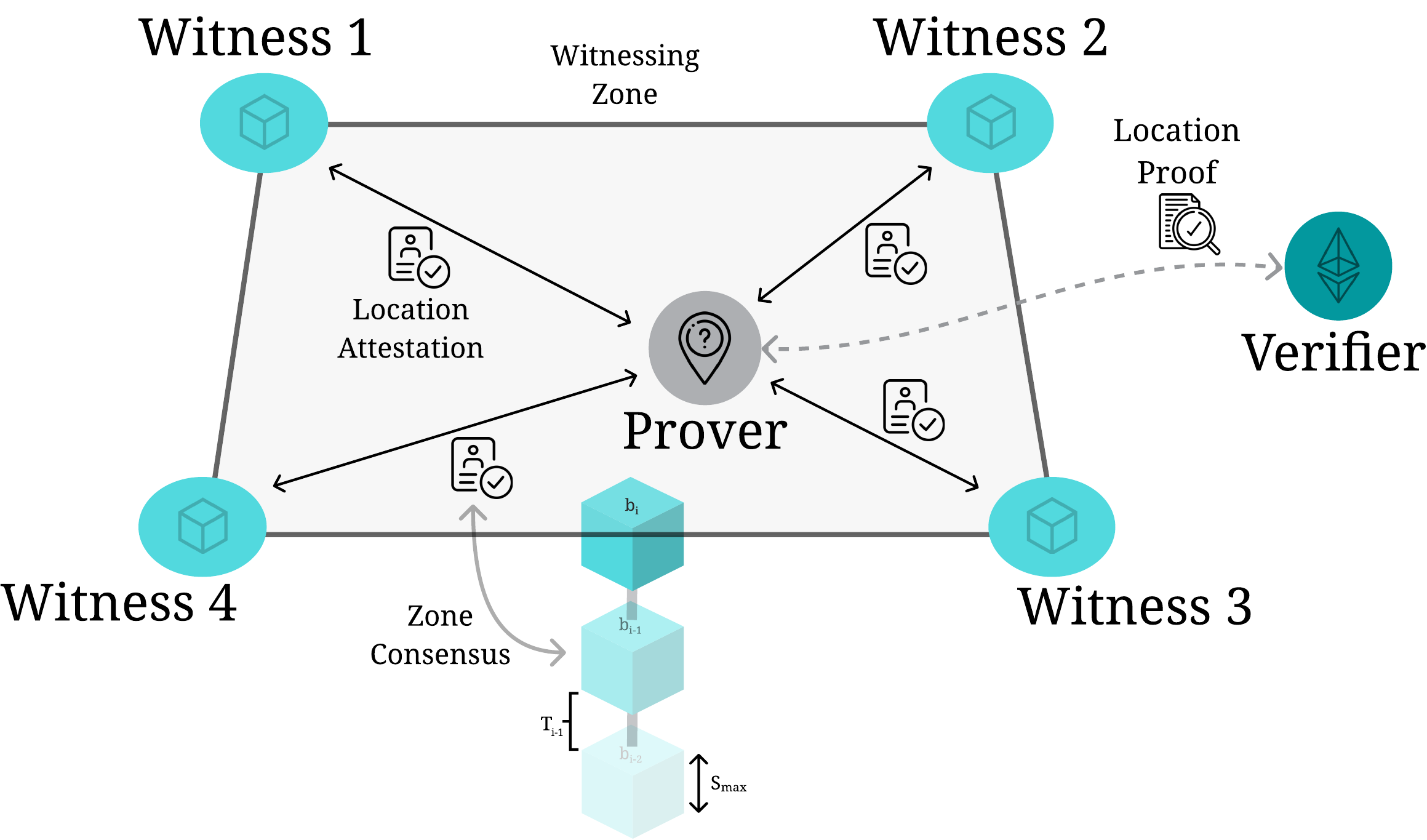}
\caption{Conceptual witnessing-zone configuration with fixed witnesses, a prover inside the zone, and an independent verifier.}
\label{fig:witnessing-zone}
\end{figure}

At a conceptual level, a witnessing zone is a bounded space-time region covered by a set of witnesses that can communicate with one another and remain logically synchronized. A separate prover-to-witness channel supports ranging and proof exchange. Figure~\ref{fig:witnessing-zone} shows the configuration used throughout this paper: the witnesses define the zone, the prover claims presence inside it, and an independent verifier later checks the assembled proof.

Trust is distributed across witnesses rather than assigned to one authority, and the model aligns with permissioned BFT protocols such as PBFT~\cite{castro1999pbft} and IBFT~\cite{moniz2020istanbul}. The unresolved question is whether the integrated system remains practical with real UWB radios, imperfect clocks, mesh links, and embedded controllers.

\section{System Model and Design}
\label{sec:system-model-design}
\subsection{Roles and Threat Model}
Building on the witnessing-zone configuration introduced in Section~\ref{sec:background-related-work}, this section fixes the concrete assumptions used by the prototype. The prototype instantiates a decentralized PoL model in which independently produced witness measurements are committed to a shared ledger and later aggregated into a verifiable proof. The three roles remain the prover, witnesses, and verifier; here we specify the claim format, fault model, and checks needed to turn witness measurements into an independently verifiable proof.

The prover announces a public identity and requests evidence of presence inside the configured zone. Each witness has known coordinates, performs local ranging, signs accepted measurements, and participates in local consensus. In the campaign, the four physical witnesses and their coordinates were administratively configured on an isolated mesh. The prototype does not yet enforce an application-level registry that binds authorized witness keys to those coordinates; witness admission is therefore an operational trust assumption. The verifier receives only a serialized proof object and checks it without access to the live witness network or the prover's own location report.

The prototype considers stale replay, one manipulated range, and out-of-zone claims. Four IBFT validators tolerate at most one faulty validator under $n \ge 3f+1$, providing ledger ordering and finality under the protocol assumptions. This bound does not by itself authenticate physical ranging witnesses or prevent unauthorized application-level claim submission. Under the campaign's closed-deployment assumption, two colluding configured witnesses could place two coordinated false ranges in a three-claim proof, leaving only one honest geometric constraint; two Byzantine validators would also exceed the ledger assumption. Denial-of-service, multi-witness collusion, unauthorized witness admission, and selective disclosure remain out of scope.

The protocol follows five logical steps: identity announcement, UWB ranging, signed claim creation, ledger commitment, and proof aggregation. A claim contains the prover identity, witness identity, measured range, timing metadata, signature, and ledger reference. A proof is accepted only when enough distinct claims satisfy the cryptographic, temporal, and geometric predicates defined below.

\subsection{Layered Architecture}
The design has four layers. The physical layer performs UWB ranging between a prover and fixed witnesses. The bridge layer reads ranging results, validates freshness, signs witness claims, and submits them to the ledger. The consensus layer maintains a permissioned BFT ledger over a local mesh. The application layer monitors the ledger, assembles claims, solves the multilateration problem, applies rejection predicates, and exports a proof for offline verification.

The ledger requires deterministic finality: once a claim is included in a finalized block, later verifiers can treat the ledger reference as a stable ordering and timestamping anchor. The mesh must expose a stable local network abstraction to the ledger while allowing the witness zone to operate without external network infrastructure such as Ethernet, Wi-Fi access points, or Internet connectivity.

\subsection{Witness-Initiated Ranging}
The abstract decentralized PoL model assumes distance-bounding exchanges between prover and witnesses. In practice, the DWM3000 modules expose packet-based UWB communication suited to DS-TWR~\cite{neirynck2016dstwr,ieee802154z}. The prover initiates the overall PoL session by announcing its identity and requesting evidence. Each witness then acts as the ranging verifier and initiates its own DS-TWR exchange by sending a poll containing a fresh challenge and ledger reference. This role split gives the witnesses control over challenge freshness and radio scheduling; it does not by itself prevent fast relay or wormhole attacks.

Each implemented DS-TWR exchange contains four frames: a witness poll, a prover response, a witness final frame, and a prover report carrying the decommitment and timing values needed for the witness to compute the range. The resulting range is forwarded to the bridge layer, which binds it to the latest local blockchain state before submitting a claim.

\subsection{Block-Hash Binding}

Stale-replay detection is addressed by embedding the latest block hash in the UWB ranging payload. A witness supplies its current block hash and a fresh challenge in the poll. The prover binds them into its commitment and returns the hash in the report, allowing the witness to check consistency. The bridge accepts only the latest or previous block hash. Because the hash is public, this check detects stale references but does not prevent fresh relay or a prover from obtaining the current hash out of band.

This mechanism complements cryptographic UWB ranging integrity mechanisms such as IEEE 802.15.4z STS mode by providing a practical application-layer freshness binding for commodity hardware and firmware.

\subsection{Multilateration and Verification}
Given witness coordinates $w_i=(x_i,y_i)$ and measured distances $d_i$, the prover estimates its position $\hat{x}=(x,y)$ by minimizing the residuals:
\begin{equation}
    \hat{x} = \arg\min_x \sum_i \left(\lVert x-w_i\rVert_2 - d_i\right)^2 .
\end{equation}
The geometric consistency of the resulting proof is summarized by the RMS residual:
\begin{equation}
    \rho_{\mathrm{RMS}}(\hat{x}) =
    \sqrt{\frac{1}{n}\sum_i \left(\lVert \hat{x}-w_i\rVert_2 - d_i\right)^2}.
\end{equation}

Acceptance requires at least three ledger-accepted witness transactions, zone containment, bounded RMS residual and timestamp spread, and membership in a triangle formed by three witnesses, following the intuition of verifiable multilateration~\cite{brito2025decentralized}. A 30 s spread bound is used during proof assembly, which accommodates prototype batching and clock recovery. Failure of any predicate causes rejection. These checks deliberately combine cryptographic, temporal, and geometric evidence, because no single layer can detect all relevant failures. The offline verifier recomputes geometry, the minimum supplied claim count, timestamps, and the prover signature. A deployment should instead derive a tighter bound from consensus latency and motion policy (e.g., $\Delta t \leq d_{\max}/v_{\max}$), and the current serialized proof also lacks complete transaction-inclusion proofs and reusable witness signatures. Real motion scenarios and ledger-independent verification remain future work.

\section{Prototype Implementation}
\label{sec:prototype-implementation}
\begin{figure}[t]
\centering
\includegraphics[width=0.72\textwidth]{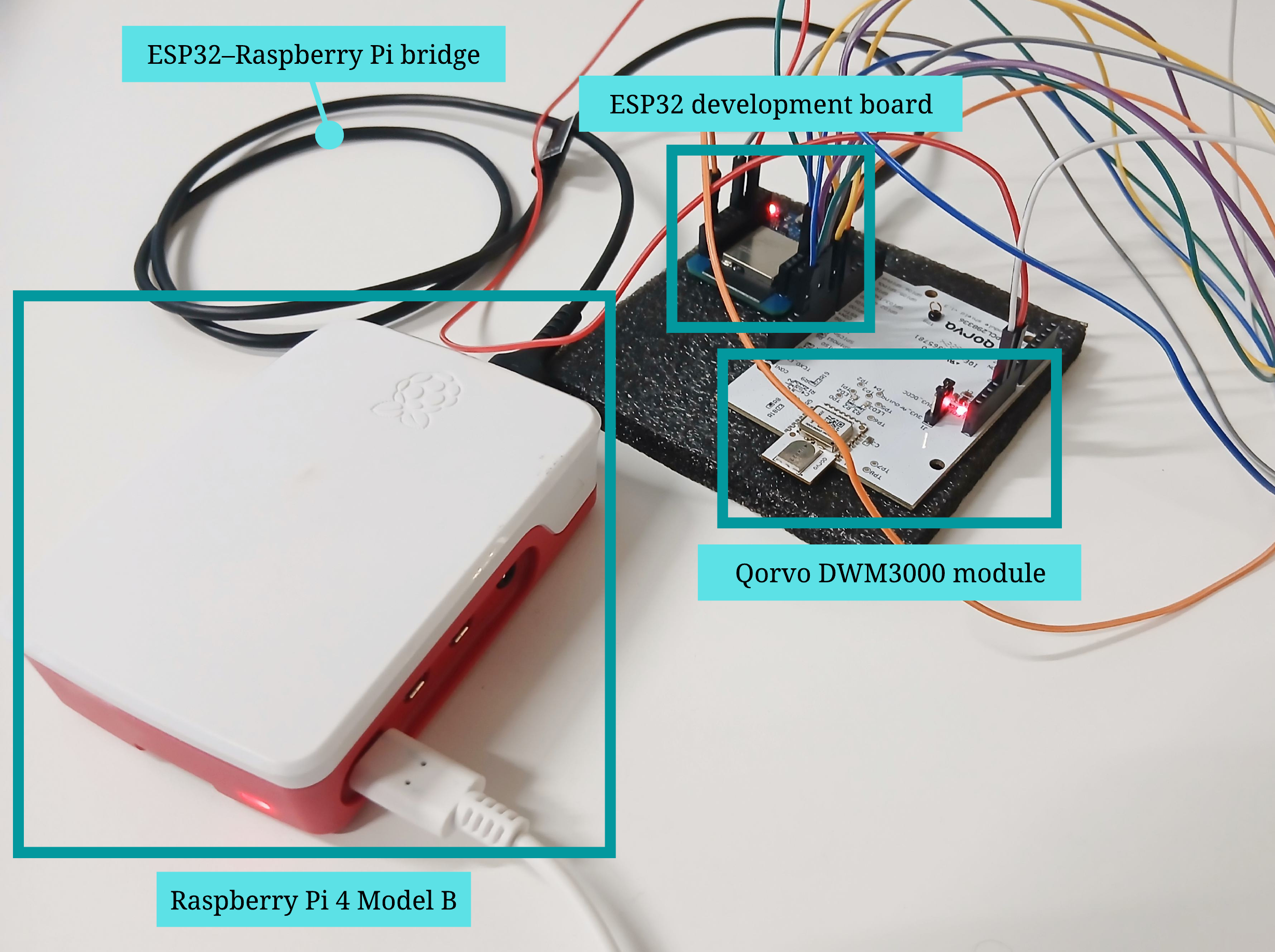}
\caption{Assembled witness node with Raspberry Pi 4, ESP32 controller, DWM3000 UWB module, and SPI wiring.}
\label{fig:witness-photo}
\end{figure}

The prototype uses five ESP32 boards equipped with Qorvo DWM3000 UWB modules: one prover and four witnesses. Each witness is controlled by a Raspberry Pi 4, as shown in Figure~\ref{fig:witness-photo}. The UWB firmware uses IEEE 802.15.4z-compatible settings on channel 5 with a 128-symbol preamble and 6.8 Mb/s data rate. The radio driver is based on an open-source ESP32-DWM3000 indoor positioning project~\cite{circuitdigest2025}, with reimplemented protocol logic for witness-initiated ranging.

The witness bridge software parses serial measurements, rejects malformed or stale records, signs distance claims, and submits them to GoQuorum. Four Raspberry Pi nodes form an isolated Layer-2 wireless mesh using B.A.T.M.A.N.-Adv~\cite{openmesh2024batman}. GoQuorum~\cite{consensys2024goquorum} runs in Docker containers with IBFT consensus~\cite{moniz2020istanbul}. The prover application collects claims, computes multilateration, applies the predicates, and serializes proofs for a separate offline verifier.

Table~\ref{tab:bom} summarizes the main hardware components. Each witness platform combines a Raspberry Pi 4 for mesh/blockchain software with an ESP32-DWM3000 pair for ranging over SPI and USB serial. The prover platform uses the same ESP32-DWM3000 hardware, runs prover firmware, and remains outside consensus.

\begin{table}[t]
\caption{Main prototype components.}
\label{tab:bom}
\centering
\small
\begin{tabular}{lcl}
\toprule
Component & Quantity & Role \\
\midrule
Raspberry Pi 4 Model B & 5 & Witness bridge / prover aggregation \\
ESP32 development board & 5 & UWB controller \\
Qorvo DWM3000 module & 5 & UWB transceiver \\
USB serial link & 5 & ESP32--Raspberry Pi bridge \\
GoQuorum validator & 4 & Permissioned ledger consensus \\
\bottomrule
\end{tabular}
\end{table}

The approximate material cost was EUR 200 per witness node, excluding ordinary cables and power supplies.

The witness bridge checks each serial record for syntax, block-hash freshness, and signing capability. Accepted records are submitted as signed transactions to a smart contract with the prover identifier, transaction sender, range, timestamp, and block reference. The contract records the sender but does not enforce a persistent allowlist or bind sender keys to configured witness coordinates; those mappings were controlled administratively during the campaign. Records that fail freshness are dropped before they can influence consensus.

The prover reads finalized claims over RPC, groups claims for the same prover in a bounded time window, and assembles the proof. A separate offline verifier receives the serialized proof object, verifies the prover's assembly signature, and recomputes the minimum supplied claim count, timestamp spread, zone containment, and residual predicates. The current proof format records witness and block metadata for traceability but does not include complete witness signatures or transaction-inclusion proofs; the verifier therefore does not yet establish an authenticated witness quorum or ledger inclusion without additional ledger evidence.

\section{Experimental Methodology}
\label{sec:experimental-methodology}
\subsection{Environment}
\begin{figure}[H]
\centering
\begin{minipage}{0.44\textwidth}
\centering
\includegraphics[width=\linewidth]{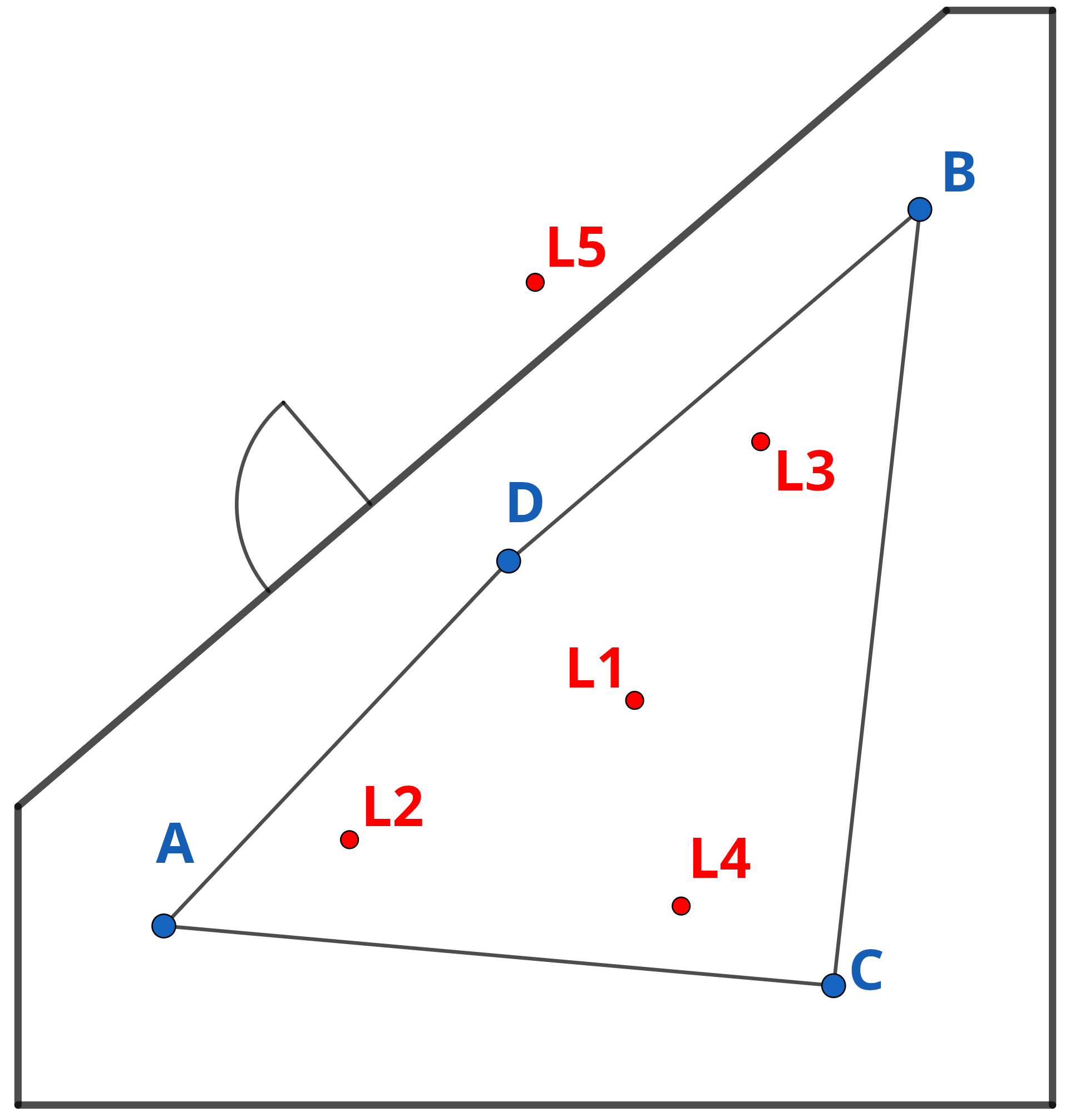}\\[-0.2em]
{\small (a) Room layout}
\end{minipage}
\hfill
\begin{minipage}{0.50\textwidth}
\centering
\includegraphics[width=\linewidth]{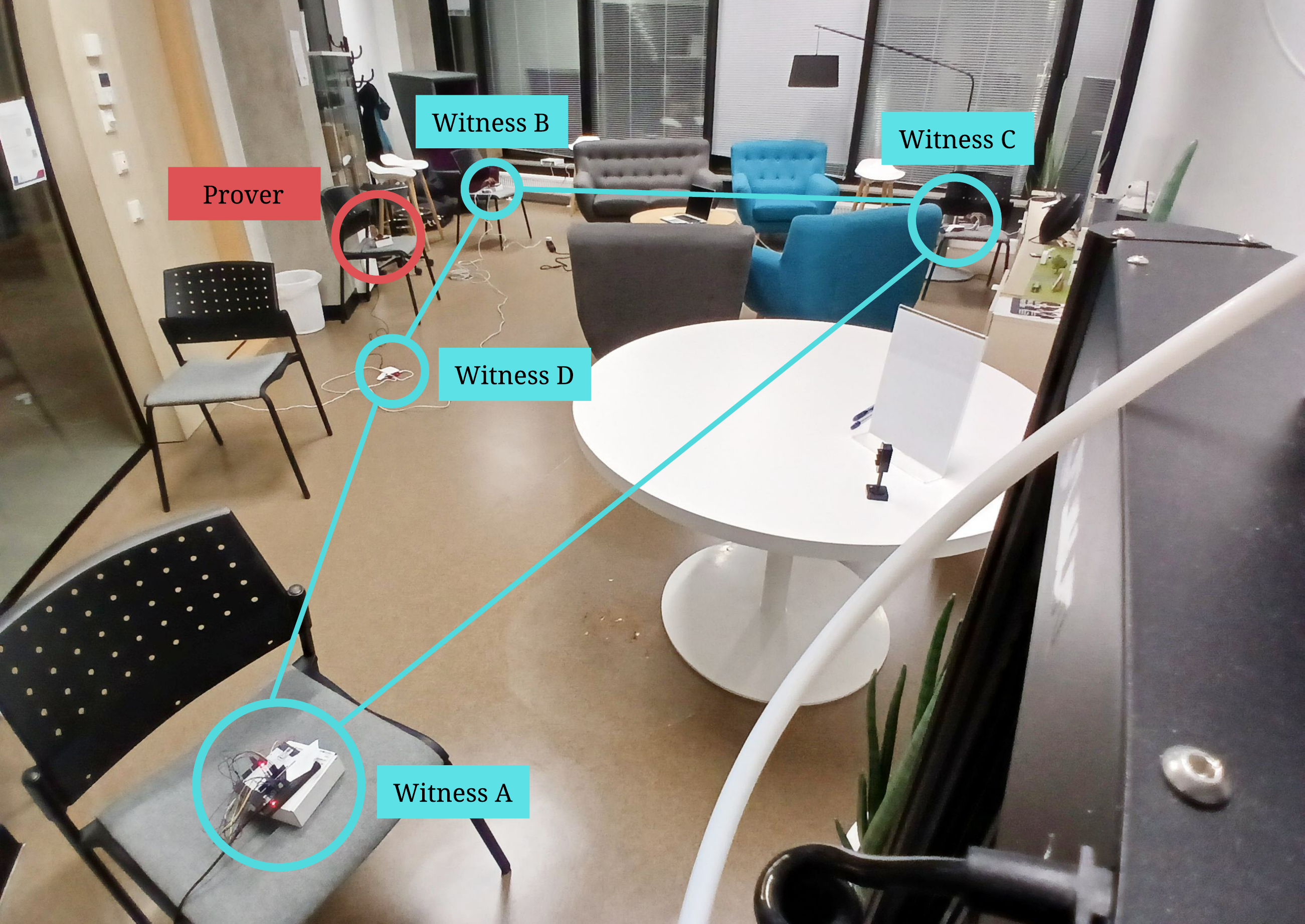}\\[-0.2em]
{\small (b) Physical setup}
\end{minipage}
\caption{Experimental environment in room 2018: measured room layout with four witnesses and five test points, and the physical setup during the campaign.}
\label{fig:experimental-environment}
\end{figure}

Experiments were conducted in room 2018 of the Delta Building at the University of Tartu. The room is an irregular pentagon with one diagonal wall. Four witnesses were placed at fixed known locations, and the prover was tested at four interior points, L1--L4. A fifth point, L5, was placed outside the configured zone as an out-of-zone control. Figure~\ref{fig:experimental-environment} shows the measured layout and physical setup.

The local coordinate frame places the origin at the bottom-left room corner. Witnesses A--D were at (1.10, 1.35), (6.80, 6.75), (6.15, 0.90), and (3.70, 4.10) m. Interior test points L1--L4 were at (4.65, 3.05), (2.50, 2.00), (5.60, 5.00), and (5.00, 1.50) m. L5 was outside the configured zone, behind a glass section of the diagonal wall, with UWB visibility and expected zone-containment rejection.

\subsection{Scenarios}
The campaign contains four phases. Phase 1 evaluates nominal operation at the four interior points and the out-of-zone control. Phase 2 stresses temporal alignment by delaying two witnesses by ten seconds while the prover is moved by approximately two meters. Phase 3 simulates a malicious witness by adding 4 m to one reported range. Phase 4 evaluates replay defense by forcing one witness bridge to inject a block hash from ten blocks in the past.

Each interior point was tested with 20 trials per phase. L5 was tested for 20 trials under Phase 1 conditions. The complete campaign therefore contains 340 proof attempts.

The scenarios exercise distinct layers. The out-of-zone control is geometrically consistent but spatially invalid. Mobility stresses ranging and aggregation under changing geometry, not the 30~s timestamp policy. Range inflation tests multilateration distortion; Phase~4 tests bridge rejection of stale references, not captured UWB replay.

\subsection{Metrics}
For each proof attempt, the system recorded Euclidean position error, acceptance flag,
categorical rejection reason, RMS residual, witness timestamp spread, proof latency,
and quorum status. Witness-side logs additionally recorded serial reads, valid ranging
records, replay rejections, submitted claims, and transaction failures.

The primary accuracy metric is the Euclidean distance between the estimated position
and the tape-measured ground truth:
\begin{equation}
    e_{\mathrm{pos}}=\left\|\hat{\mathbf{p}}-\mathbf{p}_{\mathrm{gt}}\right\|_2
=\sqrt{(\hat{x}-x_{\mathrm{gt}})^2+(\hat{y}-y_{\mathrm{gt}})^2}.
\end{equation}

Here, \(\hat{\mathbf{p}}=(\hat{x},\hat{y})\) is the estimated position and
\(\mathbf{p}_{\mathrm{gt}}=(x_{\mathrm{gt}},y_{\mathrm{gt}})\) is the
tape-measured ground-truth position in the two-dimensional room model. Security behavior is reported as acceptance rate
and categorical rejection reason, which captures policy failures such as the L5
out-of-zone control. Latency is measured from claim collection to proof assembly and
therefore captures the application-layer proof path. Binary outcomes are reported
with counts and two-sided 95\% Wilson intervals, which quantify sampling uncertainty but not external validity. Continuous metrics use the sample standard deviation
(SD) and empirical nearest-rank 95th percentile (P95).

\section{Results}
\label{sec:results}
\subsection{Nominal Operation}
Table~\ref{tab:nominal} summarizes nominal operation. Across 80 interior trials, position error was 0.180 m (SD 0.086; P95 0.305), RMS residual was 0.150 m (SD 0.061; P95 0.261), and latency was 0.0546 s (SD 0.0212; P95 0.0820). All 80 proofs were accepted with quorum (95\% Wilson CI: 95.4--100\%), and all claims reached the same IBFT block, giving 0.00 s timestamp spread.

\begin{table}[t]
\caption{Nominal operation results across 80 interior trials.}
\label{tab:nominal}
\centering
\small
\begin{tabular}{lccccc}
\toprule
Point & Error (m) & RMS (m) & Lat. (s) & Accept. & Quorum \\
\midrule
L1 & 0.23 & 0.11 & 0.052 & 100\% & 100\% \\
L2 & 0.10 & 0.11 & 0.055 & 100\% & 100\% \\
L3 & 0.11 & 0.13 & 0.057 & 100\% & 100\% \\
L4 & 0.29 & 0.25 & 0.055 & 100\% & 100\% \\
\midrule
All & 0.18 & 0.15 & 0.055 & 100\% & 100\% \\
\bottomrule
\end{tabular}
\end{table}

The highest error occurred at L4, which had the least favorable geometry. The L5 control produced geometrically consistent ranges with 0.335 m mean RMS (SD 0.026; P95 0.381) outside the polygon; 20/20 trials were rejected as \texttt{outside\_zone} (95\% Wilson CI: 83.9--100\%).

\subsection{Adversarial Scenarios}
Table~\ref{tab:adversarial} summarizes the adversarial findings. The initial 1.5 m RMS predicate did not directly reject the corrupted trials, whose residuals remained below 1.1 m; zone containment nevertheless rejected all 80 Phase 3 proofs. Applying a post-hoc 0.5 m RMS candidate would also reject 80/80 Phase 3 proofs directly (95\% Wilson CI: 95.4--100\%), preserve Phase 1 and 4 acceptance, and cause one mobility false rejection. Selection and evaluation used the same campaign, so 0.5 m is a candidate for this dataset.

\begin{table}[t]
\caption{Summary of adversarial and control outcomes.}
\label{tab:adversarial}
\centering
\scriptsize
\renewcommand{\arraystretch}{1.5}
\setlength{\tabcolsep}{4pt}
\begin{tabular}{@{}>{\raggedright\arraybackslash}p{0.22\linewidth}>{\raggedright\arraybackslash}p{0.18\linewidth}>{\raggedright\arraybackslash}p{0.48\linewidth}@{}}
\toprule
Scenario & Defense layer & Outcome \\
\midrule
Out-of-zone L5 & Zone check & 20/20 rejected by zone containment \\
10 s witness delay with movement & Time / geometry & Mostly accepted; one false rejection after RMS threshold re-tuning \\
4 m range inflation & Zone / RMS checks & 80/80 rejected by zone containment; 80/80 also rejected directly by post-hoc 0.5 m RMS \\
Ten-block-old reference & UWB / bridge & Stale contribution dropped before ledger submission \\
\bottomrule
\end{tabular}
\end{table}
\setlength{\tabcolsep}{6pt}
\renewcommand{\arraystretch}{1.0}

The 27 stationary mobility records had 0.115 m mean RMS (SD 0.064), while five transition-marked records reached 0.334 m (SD 0.210; maximum 0.649); overall latency was 0.0677 s (SD 0.0149). The recorded block-timestamp spread was 0.00 s in all 80 trials, so this phase did not exercise the 30 s rejection bound. At 0.5 m, 1/80 trials would be rejected: 1.25\% (95\% Wilson CI: 0.22--6.75\%).

A 4 m inflation produced 3.385 m mean error (SD 0.952) while the solver shifted the estimate and kept RMS between 0.619 and 1.089 m (mean 0.859; SD 0.188). Thus the 1.5 m RMS predicate missed every manipulated trial, although zone containment rejected them; the post-hoc 0.5 m RMS candidate would reject all directly. A single offset does not reveal the smallest detectable manipulation.

In 80 stale-reference attempts, a hash ten blocks old yielded no accepted contribution from the affected witness. The remaining witnesses produced proofs with 0.172 m mean error (SD 0.087), 0.071 m RMS (SD 0.027), and 0.0546 s latency (SD 0.0206).

\section{Discussion and Research Outlook}
\label{sec:discussion}
\paragraph{What did the prototype demonstrate?}
The integrated system completed physical ranging, ledger commitment, and separate proof checking with sub-meter mean error and low latency. This establishes engineering feasibility for one controlled indoor zone, not yet performance across rooms, severe multipath, concurrent provers, or zones.

\paragraph{What did calibration reveal?}
Per-witness noise thresholds do not necessarily transfer to multilateration: the 1.5 m RMS predicate missed the 4 m inflation even though zone containment rejected the resulting positions, while the post-hoc 0.5 m candidate provided direct RMS rejection with one mobility false rejection. Deployers should collect benign residuals across representative geometry and motion, choose an upper quantile plus an application-specific margin, then validate on held-out benign data and an offset sweep. This campaign performs the exploratory first step.

\paragraph{Which tested failures were exposed?}
Different layers rejected the old reference, range inflation, and out-of-zone claims, but the campaign did not test captured-transcript replay, an RMS attack boundary, or fresh relay. IBFT's one-fault bound concerns validator consensus; four configured witnesses still do not tolerate two colluders.

PoL should be treated as an evidence service with an explicit trust boundary. The prototype separates live commitment from later policy checking, but verifier independence additionally requires registered witness keys bound to coordinates, complete witness signatures, and ledger-inclusion evidence rather than trusted prover-supplied fields.

\subsection{Research Outlook}
\label{sec:limitations-future-work}
First, mobility exposed occasional UWB pipeline timeouts, likely caused by interaction between the static TDMA startup schedule and the DS-TWR state machine. This opens work on adaptive witness scheduling, negotiated ranging slots, and IEEE 802.15.4z STS-based authenticated ranging. A stronger ranging mode would also reduce reliance on the public block-hash freshness check by moving integrity closer to the radio protocol.

Second, the isolated mesh has no public NTP access, making witness clock recovery after power loss operationally fragile. Future research should compare self-contained time sources, GPS-disciplined oscillators, and consensus-derived logical clocks. The trade-off is between operational simplicity, hardware cost, and the symmetry of the witness zone.

Third, the system supports one prover and one zone; multiple provers or overlapping zones require explicit UWB scheduling. The four-validator IBFT network tolerates one faulty validator; two-fault consensus tolerance requires at least seven validators under $n \ge 3f+1$. This ledger guarantee is separate from physical-witness membership, which the prototype controls administratively rather than through persistent registered keys. Operational deployment requires contract-level admission control and verifier-side binding of distinct authorized witness identities to coordinates.

Fourth, the current proof reveals the prover address, witness coordinates, and exact distance vector. This is acceptable for a laboratory prototype. Privacy-sensitive deployments need selective disclosure, for example through zero knowledge, so that a prover can demonstrate zone membership without revealing exact coordinates or witness distances~\cite{wu2020zkpol,bogdanov2025zkpol}. A promising direction is to verify policy-level predicates, such as presence in an authorized area, without disclosing the underlying measurement vector.

Finally, one room, prover, four witnesses, and four interior positions cannot establish external validity for other geometry, materials, interference, concurrent users, or handovers. Larger, outdoor, multipath-heavy, and longitudinal deployments should study these factors alongside cost, accuracy, maintenance, churn, and validator governance.

\section{Artifact Availability}
\label{sec:artifact-availability}
The artifact\footnote{\url{https://doi.org/10.5281/zenodo.21789651}} contains the prototype implementation, source workbook, normalized data for all 340 trials, deployment output snapshots, checksums, and standard-library scripts that reproduce the descriptive statistics and Wilson intervals. It documents the workbook-to-paper phase mapping and treatment of mobility-transition records. The archived snapshot corresponds to source revision \texttt{3892b70} and is publicly available under the GPLv3 license.

\section{Conclusion}
\label{sec:conclusion}

This paper presented and evaluated a physical implementation of a decentralized Proof-of-Location witnessing zone. By integrating UWB ranging, local witness coordination, ledger-backed claim storage, proof construction, and offline checking on commodity hardware, the study shows how a conceptual architecture can be realized as an end-to-end experimental system. The evaluation demonstrates its practical feasibility in a controlled indoor environment while highlighting the interaction between physical measurements, networking, consensus, and verification.

Future work can build on this foundation through stronger ranging authentication, deployment-specific calibration, more complete verification evidence, privacy-preserving proofs, and evaluation across multiple environments, provers, and witnessing zones. More broadly, the prototype and experimental methodology provide a basis for continued research on trustworthy decentralized location evidence under real physical, networking, and timing constraints.

\paragraph{Disclosure of Interests.}
The authors have no competing interests to declare that are relevant to the content of this article.

\bibliographystyle{splncs04}
\bibliography{references}

\end{document}